# Artificial Intelligence-Defined 5G Radio Access Networks

Miao Yao[1], Munawwar Sohul[1], Vuk Marojevic[2], and Jeffrey H. Reed[1]
[1]Department of Electrical and Computer Engineering, Virginia Tech, Blacksburg, VA 24060 USA
[2]Department of Electrical and Computer Engineering, Mississippi State University, Starkville, MS 39759 USA

Corresponding author: Jeffrey H. Reed (e-mail: reedjh@vt.edu).

**ABSTRACT** Massive multiple-input multiple-output antenna systems, millimeter wave communications, and ultra-dense networks have been widely perceived as the three key enablers that facilitate the development and deployment of 5G systems. This article discusses the intelligent agent that combines sensing, learning, and optimizing to facilitate these enablers. We present a flexible, rapidly deployable, and cross-layer artificial intelligence (AI)-based framework to enable the imminent and future demands on 5G and beyond. We present example AI-enabled 5G use cases that accommodate important 5G-specific capabilities and discuss the value of AI for enabling network evolution.

## I. Introduction

Does 5G cellular communications technology in the age of intelligence really look like the Thomas W. Lawson Schooner (the last of the large cargo sailing ships) of modern times? 5G networks are anticipated to achieve system capacity improvements through aggressively increased spectral efficiency, channel bandwidth and higher densification [1]. However, concerns are raised whether this is a revolutionary leap from today's wireless communications or a simple piling up of less innovative wireless functionalities. The International Telecommunication Union (ITU) classifies 5G into three categories of usage scenarios: enhanced mobile broadband (eMBB), massive machine-type communication (mMTC), and ultra-reliable and low latency communication (URLLC) to account for more diverse services and resource-hungry applications. eMBB is a service category that addresses bandwidth-hungry applications, such as massive video streaming and virtual/augmented reality (VR/AR). mMTC is a service category enabling massive sensing, monitoring, and metering in support of the massive deployments of the Internet of Things (IoT). URLLC is a service category that supports latency sensitive services including autonomous driving, drones and the tactile Internet. Whereas 5G seems to offer virtually any service, the importance of cognitive resource management cannot be underestimated and artificial intelligence (AI)-defined 5G radio access networks are proposed to support those unprecedented requirements and leverage the emergence of mobile edge computing and caching, smart cities, and context-aware networking.

Shaped by the Turing machine theory in the 1930s and rekindled again by the advent of deep neural networks, AI approaches have been extensively applied in a wide range of research fields, including natural language processing, computer vision, and wireless communications [2]. The artificial neural network (ANN) is one of the oldest approaches to create AI. It mimics brain in its ability of massive parallelism, distributed representation, and computation. Among various types of neural networks, recurrent neural networks (RNNs) allow neurons to create and process memories of arbitrary sequences of input patterns, where the connections between layers form a loop [3]. A deep neural network or deep belief network (DBN) employs a hierarchical structure with multiple restricted Boltzmann machines (RBMs) and works through a successive learning process, layer-by-layer. RBMs form an undirected graphical model without visible-visible and hidden-hidden connections. The advantages of DBN lie in the ability of unsupervised learning, fast inference and flexibility due to its multi-layer structure. A convolutional neural network (CNN) is built on layers of convolving trainable filters that result in a hierarchy of increasingly complex features [4]. A CNN is fundamentally different from a DBN; the DBN behaves as a generative model which describes the joint distribution of data and the corresponding targets, whereas a CNN is a discriminative model which describes the distribution of targets conditioned on data. A special interest of various types of neural networks in wireless communications arises from the fact that they can perform nonlinear approximations and are intrinsically data-driven.

Recent breakthroughs in computing and AI have encouraged researchers in the field of wireless communications to leverage AI especially in the 5G context, which is illustrated in Fig. 1. An AI-defined 5G network provides the base stations (BSs)/Cloud with the capability to create a cognitive and comprehensive data repository by splitting, processing and interpreting the operational data. (The BS is defined in this article as the remote radio head plus the centralized and virtualized baseband unit pool of a single mobile network operator.) Massive amounts of real-time data are generated across a large number of users and ranges from channel state information (CSI) to IoT device readings. The received data and geolocation databases are fused to derive a complete understanding of the environment. From the human-centric communication perspective, the human behaviors are learned and adapted by the reconfigurable AI-defined wireless network to evolve the network functionalities and create people-oriented services. From a machine-centric communication perspective, big data analytics are leveraged to extract massive patterns, especially at the physical (PHY) and medium access control (MAC) layers and enable self-organizing operations. Neural networks can be used to redefine communication networks, solving a number of nontrivial design problems at runtime and across layers for cognitive link adaptation, resource scheduling, signal classification, and carrier sensing/collision detection, among others. An RNN has also the capability to capture and mitigate the imperfections and nonlinearities of radio frequency (RF) components, such as high-power amplifiers (HPAs), which incur at the PHY and can affect the network performance. DBN and CNN are better suited for resolving a range of upper communication layer tasks such as network optimization and resource management. This article briefly examines the key 5G enablers and use cases and what role AI may play. We discuss some of the emerging PHY and MAC layer issues of 5G networks and propose two AI-based 5G radio access technology (RAT) architectures and illustrate their potentials through numerical results. We conclude the paper with pointing out some limitations, open issues and research challenges.

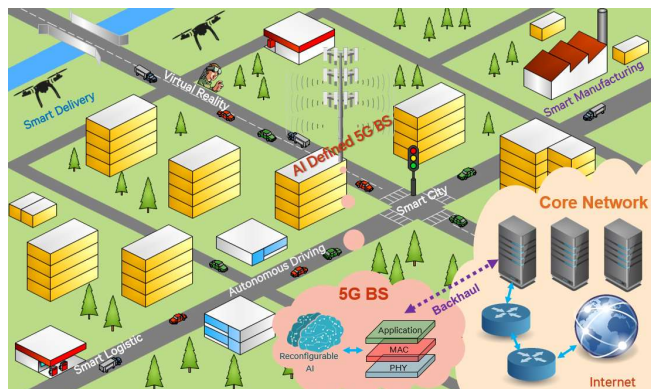

| 5G Services Type | Applications |
|---|---|
| Broadband | VR/AR |
|  | Massive Streaming |
| Distributed | Mobile Cloud Computing |
|  | Smart Infrastructure |
| Omnipresent | Massive IoT |

**FIGURE 1.** Application scenarios of AI-defined 5G networks.

## II. Artificial Intelligence Supporting 5G Enablers

The complexity of 5G context and the power of AI allow rethinking cellular communications. Figure 2 shows the critical 5G use cases and the application of different types of neural networks as key 5G enablers. Massively parallel processing devices such as graphic processing units (GPUs) and neural network processing units (NPUs), which can be deployed in a 5G BS/Cloud, enable offloading the traditional data processing and have shown to be highly energy-efficient for neural network learning and inference [5]. This section discusses the benefits of an AI framework to satisfy the technical challenges of 5G.

### Massive MIMO

Massive MIMO is widely perceived as a leading candidate technology for 5G. The high number of BS antennas requires a large number of PAs, one per antenna. The primary problem in RF PAs is known as the design trade-off between linearity and efficiency: amplifiers can be designed to attain good linearity at the cost of efficiency [6]. While the highly linear PAs are expensive and power-inefficient, the excessive number of BS antennas makes the use of inexpensive components desirable to keep the overall capital expenditures (CAPEXs) and operational expenditures (OPEXs) manageable [1]. On the other hand, emerging energy- and spectrum-efficient wideband wireless communication systems are vulnerable to nonlinear distortions that are attributed to the RF front ends, especially those of the HPAs, which affect the performance of the intended receiver and the entire network.

Furthermore, as next generation communication systems will operate in more dynamic environments and in different bands, the dynamic range requirements will likely become more demanding. Therefore, the PAs need to meet stricter linearity specifications while maintaining an acceptable efficiency. The highest PA efficiency is achieved when constantly feeding the PA, at the limit of its highest-power linear region. This is not a feasible solution for high peak-to-average power ratio (PAPR) signals and is not realistic for 5G BSs. Instead, signal processing-based solutions are used to provide a better cost-performance trade-off.

One of the main advantages of ANNs is their ability to extract, predict, and characterize nonlinearities from massive datasets. Since ANNs can capture many kinds of relationships between a variety of time-varying inputs and outputs, such as complex nonlinear relationships or highly dynamic relationships, they are well suited to tackle nonlinearity problems at the PHY, including PA nonlinearity tracking, predistortion and impairment correction. A detail discussion of AI-aided energy-efficient beamforming is provided later.

### Extended RF Spectrum and Spectrum Management

The demand for higher end-user data rates and higher traffic density drive 5G networks to provide a corresponding capacity boost. Millimeter wave (mmWave) systems have access to a huge amount of available bandwidth in extremely high frequency bands, between 30 and 300 GHz, to enable multi-gigabit-per-second rate connectivity. Previously, mmWave at 60 GHz band has been adopted by local and personal area communication standards such as IEEE 802.11ad and IEEE 802.15.3c, as well as for certain small cell backhaul deployments. mmWave communications are facilitated by highly directional communication links, where narrow beams can focus the energy to the intended receivers, reduce PA requirements and suppress interference from other transmitters. Whereas the mmWave spectrum may not be scarce, it is also not very reliable due to its propagation characteristics. At mmWave frequencies severe channel variations occur due to blockage by cars, trees, solid building material, and even human body parts.

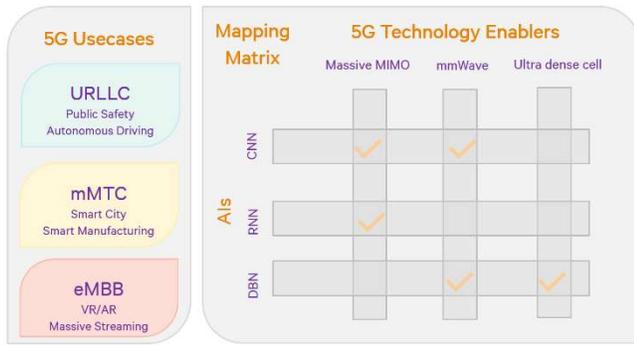

FIGURE 2. 5G use cases, key technology enablers and typically applicable AI principles (CNN: convolutional neural network, RNN: recurrent neural network, DBN: deep belief network).

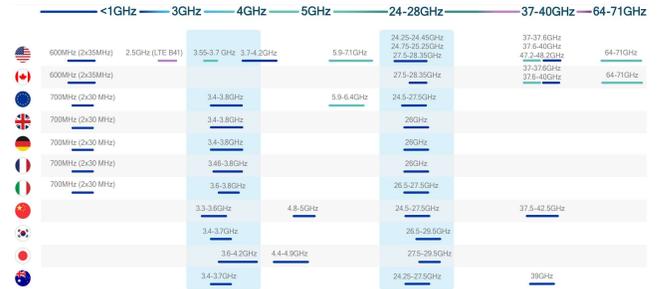

FIGURE 3. Global snapshot of 5G spectrum (blue: Licensed, green: unlicensed/shared, purple: existing bands) [7].

As a result, a hybrid of spectrum landscape of low and high frequencies necessary to maintain seamless network connectivity and enable the 5G verticals [7]. This hybrid spectrum includes micro and mmWave bands and different types of licensing (Fig. 3). AI allows smart scheduling and cooperation of different spectrum portions since learning and inferencing can be carried out based on user behaviors and network conditions. Specifically, neural networks can provide multi-RAT BSs with the capability for real-time resource management. They also enable self-planning, self-organizing, and self-healing, where a group of distributed and autonomous AI agents are distributed among different BSs. An example of an AI-enabled spectrum access system (SAS) that manages multiple RATs in shared spectrum is discussed in later sections.

**Network Ultra-Densification**
It is envisioned that the explosive data traffic requirement in 5G will be largely dependent on the deployment of ultra-dense small cells [8]. The technical challenges include finer-granularity resource allocation and cross-tier interference mitigation. In fact, the cell densification is redeemed as one of the most aggressive ways to improve system capacity as long as the interference can be efficiently managed.

Co-channel interference mitigation in 5G is naturally enabled through massive MIMO beamforming and mmWave transmission. When combined with a centralized AI-aided learning, the envisioned 5G network will benefit from both AI and small cell deployments, where complex radio resource management at "pixel" level can be performed more efficiently. For instance, the radio environment map (REM) is promising to provide a practical method for the realization of AI-enabled interference mitigation based on dense geolocation databases. It builds long-term knowledge by collecting historical radio environment measurements and radio scenes. The radio network prediction, evolution and optimization can be made by active learning based on network coverage, capacity and quality maps.

The AI agent in the baseband unit (BBU) pool assists with the baseband processing functions, such as the radio resource management, interference management, and handover control. Such AI-driven self-organizing solutions provide more opportunities for design optimization of self-organized networking, increase the level of flexibility and achieve cost reductions. Moreover, the centralized processing of CSI by the AI agent allows additional insights into the channel that perform closer to channel capacity and schedulers that provide better fairness and end-user experience.

### III. Artificial Intelligence for Diverse 5G Application Scenarios and Services

To further elaborate on the importance of the revolutionary 5G paradigms wherein AIs play critical roles, this section discusses a few disruptive use cases of 5G. These applications have stringent requirements on different communication layers; here we focus on the 5G network implications.

**Vehicular Networks**
Vehicular networks leverage wireless technologies to improve traffic safety and transportation efficiency, which is specifically useful for active driving assistance. It involves a broad range of stakeholders such as vehicle manufacturer, wireless policy regulators and transportation authorities. A myriad of vehicle applications is envisaged, including automated toll collection systems, autonomous driving systems, and information provisioning systems. Autonomous driving systems are capable of detecting changes in the surrounding by combining sensor data with signal and image processing to inform the driving system of hazardous road conditions. For example, the real-time wet road condition detection and anti-collision detection are necessary to improve not only vehicular safety, but also efficiency.

The recent advances in 5G networks and AI enhance the vehicular networks by provisioning reliable, efficient, low-latency, and cooperative communications. The autonomous vehicles collect the vehicle-centric information (e.g. speed, global positioning, device functionality, and performance), road-centric information (e.g. road surface condition, routing, collision) and passenger-centric information (e.g. drowsiness, length of continuous driving) and send it to the AI-defined 5G BS. The BS is capable of learning the behaviors of autonomous vehicles on the road, predicting the speed of vehicles, driving states, obstacles and lane-changing behaviors to assist transportation systems with processed information and improve road mechanics.

### VR and AR
VR and AR over wireless are expected to become the 5G killer applications. Information is perceived through audio, visual, and touch in VR. Computer Generated (CG) content is combined with real objects and visualized in AR. VR is a key use case where both URLLC and eMBB play important roles to (1) avoid motion sickness introduced by motion and visual coordination latency and (2) facilitate the necessary capacity increase that ultra-high-definition video streaming requires. In particular, today's VR systems experience both computing and communication delay bottlenecks. Moreover, heavy image processing tasks require high computing power which is often not available locally. In emerging 5G networks, computing resources at the BSs could serve as the VR control center and connect to the VR devices over the wireless network. Dense 5G network nodes offer the opportunity of decoupling image recognition and rendering from devices, where the major computing tasks can be done at the network edge to relieve the computing and power burden on the users' devices. However, this design approach introduces rate-latency-reliability tradeoffs. In addition, the BSs need to collect sensor information from the VR devices to optimize its processing and provide interactive and immersive user experiences. Several users may be physically coupled to perform collaborative tasks in a virtual or semi-virtual world. Hence, massive user data, including sensor data, CSI, battery life, geolocation, and user gestures need to be collected by the network for AI-aided resource scheduling and big data processing. AI-defined 5G networks enable balancing the in-VR/AR vs. in-network computing and adaptively make use of powerful GPUs at BSs or other edge nodes in an on-demand fashion. This allows trading the rather scarce edge resources against the abundant Cloud computing resources as a function of scenario and application.

## IV.  AI-Defined 5G RAT: PHY and MAC
The ever-increasing demand for energy-efficient and rich-content applications, requires a paradigm shift in 5G. In practice, a number of spectrum distorting effects such as in-phase and quadrature (IQ) imbalance and nonlinear distortion introduce out-of-band emission and energy efficiency losses. Two critical components of AI-defined network

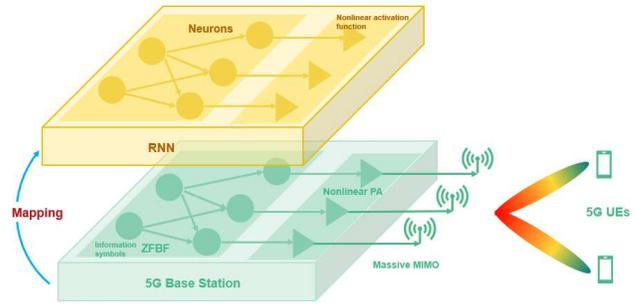

FIGURE 4.  Zero-forcing beamforming in 5G BSs: followed by nonlinear PAs (recall that RNNs are composed by interconnected neurons followed by nonlinear activation functions).

enhancements are introduced in this section to facilitate effective resource management.

### RNN for the PHY: AI-aided energy-efficient massive MIMO beamforming
AI techniques have been investigated for wireless communication systems to efficiently mitigate nonlinearities at the PHY [9]. In particular, RNNs lend themselves well to address RF nonlinearity issues because they allow historical information to be stored through the loop-based structure. These networks were originally applied to solve linear programming problems and inspired various researches to address nonlinear problems. RNNs are capable of automatically extracting nonlinear system features that are invariant to the environment and can outperform traditional approaches in PHY signal processing [10].

As discussed earlier, energy efficiency of a BS significantly depends on the PA efficiency. The deployment of massive MIMO systems, where BSs are equipped with a significantly larger number of antennas than the number of simultaneously served users, creates excessive degrees-of-freedom (DoFs) [1]. The level of underdetermination of an underdetermined channel matrix increases with the number of BS antennas when assuming the number of users is known. The excessive DoFs enable selecting the transmit signals from a larger solution space. The objective of energy-efficient beamforming in a massive MIMO system is then to find the beamforming matrix which introduces minimum PA nonlinearities within this solution space. This enables the use of low-cost PAs and their operation at higher power levels for increased energy efficiency.

Consider a downlink massive MIMO-OFDM system with $M_r$ single antenna users, one BS equipped with $N_t$ antennas, and $N_c$ subcarriers [10]. The number of BS antennas is significantly larger than number of users, i.e. $N_t \gg M_r$. Zero forcing (ZF) beamforming is applied and user information symbols at the BS are mapped to the appropriate transmit antenna to ensure minimal interference. The excessive DoFs allow choosing a solution that maximizes energy efficiency and minimizes the effects of PA nonlinearity.

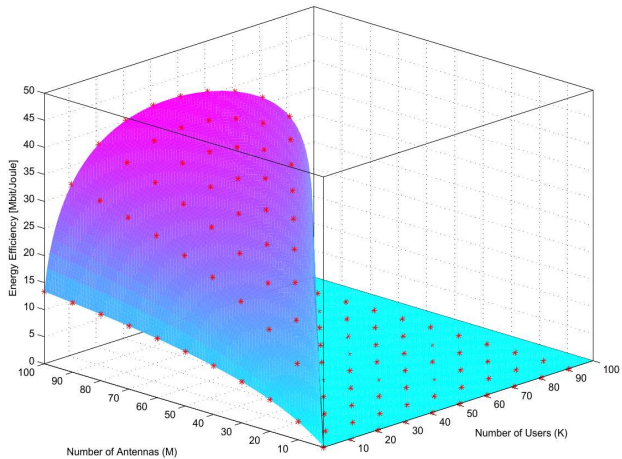

**FIGURE 5.** Energy efficiency (in Mbit/Joule) with ZF beamforming in RNN-aided massive MIMO system.

An RNN-based solution is inspired by the similarity between beamforming-PA architecture in MU-MIMO systems and the neuron-activation function architecture in RNNs, as illustrated in Fig. 4. The intuition of the neuron weights training behind this solution is as follows: The nonlinear activation function in the RNN learns about the nonlinearities of the PAs recursively and finds suitable neuron weights that satisfy two requirements: 1) zero forcing beamforming, which means minimal MU interference and 2) minimum overall nonlinear distortion. The RNN models the nonlinearity of the PA array and it is optimized to minimal transmitted power while providing a ZF solution. The RNN then informs the 5G system on how to set its weights of ZF beamforming.

The RNN-based scheme is simulated for a massive MU-MIMO-OFDM downlink system with 128 subcarriers, up to 100 antennas at the BS and up to 100 single-antenna users. The optimal transmitted signal and activation variable are derived. The relationship between the antenna configuration and energy efficiency is illustrated in Fig. 5. The result shows that huge energy-efficiency improvements can be achieved with this scheme as long as there is sufficient excess in the DoFs [10]. This also suggests that it is efficient to deploy more (low-cost) antennas as their CAPEX will be quickly overcome with a reduction of OPEX.

## DBN for the MAC: AI-Aided Spectrum Access System

The spectrum access system (SAS) will be a component of 5G networks that is expected for systems operating in the sub-6 GHz band, the 24-71 GHz mmWave bands, as well as in unlicensed bands (Fig. 3). Distributed sensing nodes in the mmWave band will likely be deployed to help gathering and extracting meaningful information about the channel quality and use while providing massive datasets to the SAS. These sensing nodes, along with additional signaling and a-priori information, provide enhanced capabilities to the SAS which will further process the contextual information to infer and learn how to best allocate spectrum.

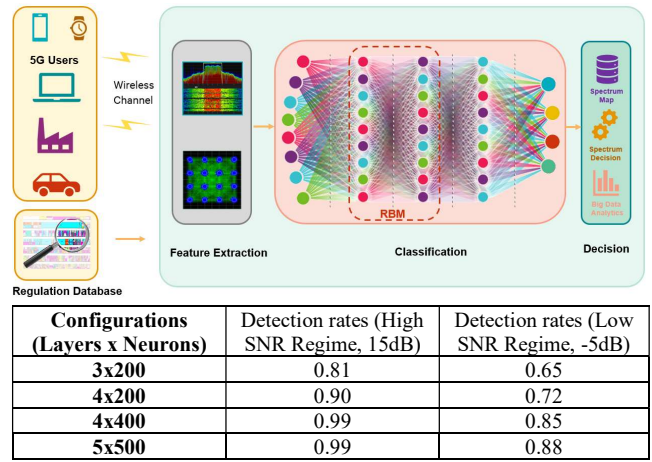

| Configurations (Layers x Neurons) | Detection rates (High SNR Regime, 15dB) | Detection rates (Low SNR Regime, -5dB) |
|---|---|---|
| 3x200 | 0.81 | 0.65 |
| 4x200 | 0.90 | 0.72 |
| 4x400 | 0.99 | 0.85 |
| 5x500 | 0.99 | 0.88 |

**FIGURE 6.** Proposed SAS architecture in 5G networks and performance comparisons (Detection accuracy of 20 MHz 802.11ac signal and LTE signal which are translated to 3.5 GHz shared band).

Previously, we demonstrated our spectrum sharing contribution for next generation public safety networks [11]. We analyzed the feasibility of broadband public safety applications in the 3.5 GHz band using a cognitive repeater. The proposed system architecture provides flexibility for storing operational data, which reduces the reporting overhead. In this article we enhance the SAS with an AI-aided framework to facilitate intelligent spectrum regulation in 5G by training it with previously generated operational data and REMs.

A reconfigurable deep learning framework based on our proposed AI-aided 5G BS system is shown in Fig. 6 for interpreting the signaling environment. It includes an offline training phase (including training and fine-tuning) and an online spectrum access phase. This framework leverages the fact that the trained model (especially the weights) can be efficiently stored to facilitate online spectrum access decisions in real time. It aims to extract meaningful spectrum usage information from massive spectrum datasets and regulatory databases in different radio contexts and provide comprehensive baseline knowledge of efficient spectrum access solutions. With the capability of distributed spectrum monitoring and awareness, the centralized SAS in the 5G BS can improve the spectrum utilization. To demonstrate this framework, we use a stack of RBMs to train the DBN in a layer-by-layer manner to reduce the training complexity. The network first models the input datasets of spectrum and regulation with the parameters of the first layer of RBM. It then trains the subsequent layers of RBM with the samples from the output of the previous layer. The contrastive divergence of the 1-step iteration (CD-1) method is applied to update the weights in each layer [3] thus allowing to find the parameters for all layers of the RBM. In the online spectrum access phase, a spectrum decision is made to generate proper transmission schemes, that is, a suitable access method, carrier frequency and bandwidth, modulation and coding scheme, and transmission power.

The simulation results for the interference recognition in the SAS are shown in Fig. 6, which demonstrate how DBN is applied to detect the signals in the 3.5 GHz band in high and low signal to noise ratio (SNR) regimes. DBNs are trained with frequency domain data. The payload data of 802.11ac and physical downlink shared channel (PDSCH) of LTE are both QPSK modulated. The results show that the DBN-based signal detection achieves a very high detection rate in the high SNR regime with 4 layers and 400 neurons in each layer.

## V. Limitations, Challenges and Open Research Problems

The success of AI-defined 5G networks relies on solving a number of research challenges across networks of heterogeneous capabilities and different levels of context-awareness. Technical issues such as network function virtualization, environmental awareness and security challenges, among others, are expected to have a significant impact on ongoing AI-defined 5G network research. Some of the critical limitations, challenges and open research problems are discussed in continuation.

### AI at-the-edge or AI over-the-bridge?

The developments of mobile edge computing and caching make it possible for mobile devices to satisfy the demands of intelligence locally. Enormous types of xPUs which are implementations of AI hardware accelerators have been introduced over the past few years. (The different types of xPUs include the brain processing unit, the deep learning processing unit, the intelligence processing unit, and the NPU.) Scalable and distributed AI agents that are distributed across a plethora of devices along with appropriate training data. This approach is referred to as on-device AI or *AI at-the-edge*. The AI-powered smartphones with limited application context such as AI-assisted photography or Apple's Memoji have been extensively adopted by smartphone manufacturers in recent years and this concept is likely to be adopted for physical and network layer problems. There are various advantages for AI at-the-edge including the available features of geographical location services (GPS, cellular triangulation), connectivity (4G/5G, Wi-Fi), and many types of non-standard sensors (gyroscope, barometer, accelerometer). However, the computational power, memory, and battery life present potential obstacles to intensive 5G services. As a result, it may be difficult to implement large-scale, complex AI-defined 5G functionalities due to the limited resources that are available in current generation smartphones, although this may change in the future.

AI applied at the BS/Cloud, the so-called *AI over-the-bridge*, has access to considerable computing resources, memory bandwidth and power supply, which are all necessary for supporting 5G services with vastly heterogeneous requirements through efficient resource scheduling and co-ordination. AI over-the-bridge overcomes the resource shortcomings with massive resource availability. Models that are trained on the device for later AI over-the-bridge deployment could reduce the uplink data traffic. As a result, the AI-defined 5G networks should leverage the available edge resources, learn and predict the user needs and environmental circumstances and periodically cache the pre-trained models from edge users to the network for subsequent use.

### Channel State Information

Accurate channel estimation is necessary at the receiver for coherent detection and at the transmitter for beamforming and power control. The CSI is one of the most important factors for inference, prediction, and big data analytics in AI-defined 5G systems. The rich context information and large amount of CSI pose significant potential for facilitating novel AI-enabled applications beyond immediate context awareness. With massive CSI, AI enables 5G networks to learn and adapt from the datasets generated by the users and BSs in different environments.

The overhead of channel estimation pilots in the frequency division duplex (FDD) mode scales linearly with the number of transmit antennas, which limits the practical number of antennas. When operating in time division duplex (TDD) mode, the massive MIMO-equipped 5G BS obtains the uplink CSI via uplink pilots, and downlink CSI is implicitly obtained due to channel reciprocity. However, the AI models for downlink usage which were trained by uplink CSI might be misleading in some cases. One reason is that the RF components and processing chains at the BS are generally not reciprocal between the uplink and downlink, and the inexpensive hardware components in the analog domain used by massive MIMO systems can exacerbate this problem further. Inaccuracies in channel estimation caused by pilot contamination can also introduce AI model training errors. In addition, the massive volumes of CSI inherent with large-scale MIMO systems pose challenges to storage and real-time processing. Redundancies of the CSI data need to be eliminated to reduce processing latency and resource overhead and imperfect CSI estimates need to be considered in the decision-making process.

### Security and Privacy

Security and privacy are critical in wireless networks and especially for 5G networks, which will provide mission-critical services. First, AI agents themselves can be the target of attack from a malicious adversary. For example, a selfish user operating in the AI-defined SAS context could broadcast fake signals to the BS and users to manipulate the AI training so that spectrum is denied to other users, which enabling the malicious party to gain access to a large portion of the shared channels. In addition, attacks through the wireless channel, such as denial-of-service, spoofing, and malicious data injection, could also disrupt the AI agent and lead to misbehaviors. Since massive amount of training datasets are required for training AI, efficient and expeditious detection of malicious training contents becomes challenging. Therefore, collaborative efforts among various parties and stakeholders are required to prevent security breaches and privacy leaks and ensure the trustworthiness of data and decisions.

## VI. Conclusions

In this article we introduced the concept of AI-defined 5G networks and discussed how AI agents entail some of the key 5G enablers and critical application scenarios. We presented AI-aided applications and examples for solving different aspects of 5G network deployment and management and discussed two architectures for the PHY and MAC layer processing at the BS/Cloud. We also elaborated on some limitations, challenges and open issues for researchers to address. We encourage AI-defined 5G and beyond research and collaborations between AI and wireless communications scientists and practitioners, who need to share their insights and data just like the envisaged network nodes do in the age of contextual awareness and intelligence.

## Acknowledgment

This work was in part supported by the National Science Foundation through Award Number 1564148. The authors would like to thanks Rod J. Walton for his inspirational advice on this article.

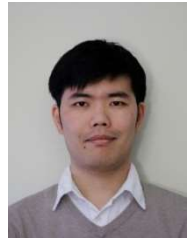

**Miao Yao** (miaoyao@vt.edu) received his B.S. degree in electrical engineering from Beijing Jiaotong University in 2007, M.S. degree in microelectronic and solid-state electronics from Tsinghua University in 2010 and Ph.D. degree in electrical engineering from Virginia Tech in 2017. In Spring 2017, he joined Qualcomm modem hardware development team. His primary research interests are dynamic-spectrum access related research issues and energy-efficient resource allocation algorithms in next generation cellular communication systems.

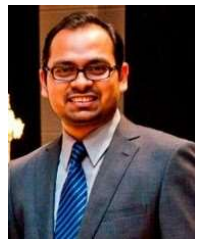

**Munawwar M. Sohul** received Ph.D. in Electrical Engineering from Virginia Tech. Currently, he is a working as Senior Engineer at Federated Wireless. His research interest in wireless communication and signal processing includes spectrum sharing, opportunistic spectrum access.

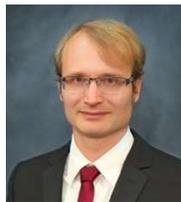

**Vuk Marojevic** (vuk.marojevic@msstate.edu) is an associate professor in electrical and computer engineering at Mississippi State University, Starkville. His research interests are in resource management, vehicle-to-everything communications, physical-layer security, spectrum sharing, software radios, and wireless network virtualization with application to commercial cellular communications, mission-critical networks, and unmanned aircraft systems.

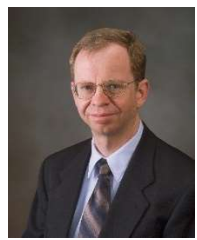

**Jeffrey H. Reed** (reedjh@vt.edu) is the Willis G. Worcester Professor in the Bradley Department of Electrical and Computer Engineering, Virginia Tech. He is the author of three books and over 200 journal and conference papers. He received his B.S., M.S., and Ph.D. degrees from the University of California, Davis.